\documentclass[aps,prl,reprint,
 superscriptaddress,
 longbibliography,
 amsfonts,amssyb,amsmath,preprintnumbers,floatfix,showpacs,
 letterpapaer,byrevtex,citeauthorscript]{revtex4-1}

\usepackage[T1]{fontenc}	
\usepackage[latin1]{inputenc}	
\usepackage[english]{babel}	
\usepackage{mathptmx}		
\usepackage{bm}			
\usepackage{microtype}		
\usepackage{mathtools}		
\usepackage{graphicx}		
\usepackage{xcolor}		
\usepackage{booktabs}		
\usepackage{comment}		
\usepackage{epsfig}
\usepackage{colordvi}
\usepackage{color}

\RequirePackage{xspace}

\providecommand{\mbf}{\mathbf}
\renewcommand*{\vec}[1]{\mbf{#1}}
\newcommand*{\A}{\vec{A}}

\begin{document}

\title{Twisted soft photon hair implants on Black Holes. }


\author{Fabrizio Tamburini}
\email{fabrizio.tamburini@gmail.com}
\affiliation{ZKM -- Centre for Art and Technologies for Media,
Lorentzstra{\ss}e~19, D-76135 Karlsruhe, Germany}
\affiliation{MSC -- bw,  Stuttgart, Nobelstr. 19, 70569 Stuttgart, Germany.}

\author{Mariafelicia De Laurentis}
\email{laurentis@th.physik.uni-frankfurt.de}
\affiliation{Institute for Theoretical Physics,
 Goethe University, Max-von-Laue-Str.~1, D-60438 Frankfurt, Germany}
\affiliation{Tomsk State Pedagogical University, ul. Kievskaya, 60, 634061 Tomsk, Russia.}
 \affiliation{Lab.Theor.Cosmology,Tomsk State University of Control Systems and Radioelectronics (TUSUR), 634050 Tomsk, Russia.}
 \affiliation{INFN Sezione  di Napoli, Compl. Univ. di Monte S. Angelo, Edificio G, Via Cinthia, I-80126, Napoli, Italy.}

\author{Ignazio Licata}
\email{iagnazio.licata@ejtp.info}
\affiliation{Institute for Scientific Methodology (ISEM) Palermo, Italy.}
\affiliation{School of Advanced International Studies on Theoretical and Nonlinear Methodologies of Physics, Bari, I-70124, Italy.}
\affiliation{International Institute for Applicable Mathematics and Information Sciences (IIAMIS), B.M. Birla Science Centre, Adarsh Nagar, Hyderabad -- 500 463, India.}

\author{Bo Thid\'e}
\email{bt@irfu.se}
\affiliation{Swedish Institute of Space Physics,
 {\AA}ngstr\"{o}m Laboratory, P.\,O.~Box~537, SE-75121, Sweden} 

\begin{abstract}

The Hawking-Perry-Strominger (HPS) work \cite{hawk16} states a new
controversial idea about the black hole (BH) information paradox
\cite{hawk74,hawk75,hawk76,page04} where BHs maximally entropize and
encode information in their event horizon area \cite{hb1,hb2}, with no
``hair'' thought to reveal information outside but  angular momentum,
mass and electric charge only \cite{wheeler,chandra} in a unique quantum
gravity (QG) vacuum state.  This new idea invokes new conservation
laws involving gravitation and electromagnetism \cite{strom,strom2},
to generate different QG vacua and preserve more information in soft
photon/graviton hair implants.  In the context of black holes and the
HPS proposal we find that BH photon hair implants can be spatially shaped
\emph{ad hoc} and encode structured and densely organized information on
the event horizon involving novel aspect in the discussion a particular
aspect of EM fields, namely the spatial information of the field
associated to its orbital angular momentum.  BHs can have ``curly'',
twisted, soft-hair implants with vorticity where structured information
is holographically encoded in an organized way in the event horizon.

\end{abstract}

\maketitle

The BH information paradox, information loss and the no-hair theorem were mainly based on the supposed unicity of the quantum gravity vacuum that revealed outside only the macroscopical quantities angular momentum, mass and electric charge.
HPS changed this scenario with new results in GR and QG, new invariants and symmetries such as supertranslations \cite{bondi,strom} and superrotations \cite{donnay} and observable effects \cite{mirba,compere,shei}. Anyway, in our knowledge, it is still an open question whether HPS soft hair can account for the Bekenstein-Hawking entropy and capture the information of black-hole microstates.

When one includes the set of conserved quantities of the electromagnetic (EM) field too \cite{strom2,fuschich,thidebook,review}, EM analogs of the supertranslation charges are introduced as a generalization of the electric charge conservation principle. 
Also in this case HPS suggest that the unicity of the QG vacuum state is invalidated because of the creation or annihilation of actual physical quantum zero-energy state particles, soft photons.
When EM phenomena act on BHs, BHs are expected to carry a soft electric hair implant, made with soft photons that store information of any process in the event horizon, with different QG vacuum states and different BHs with identical macroscopical parameters.

Consider the metric obtained from the collapse of neutral matter at advanced time $v = 0$,
\begin{equation}
ds^2=  - \left( 1- \frac{2M\Theta(v)}r \right) dv^2  + 2dvdr + 2r^2\gamma_{z \bar z} dz d \bar z,
\label{metricbh}
\end{equation}
written in terms of the round metric $\gamma_{z \bar z}$ on the unit sphere $S^2$ in complex coordinates $(v,r,z,\bar z)$,
where $v=t+r$ and $r$ the radial coordinate, $z=\tan (\phi /2) \exp(-i \theta)$ and $z=1/\bar z$, $M$ the BH mass and $\Theta=0$ before the shell at $v = 0$ and $\Theta=1$ after the shell at $v = 0$.
The complete information is stored in the $S^2$  future boundary of the horizon, $\mathcal{H}^+$ of the BH horizon, $H$, an holographic plate made of quantum pixels whose excitation corresponds to the creation of a spatially localized soft photon on the event horizon with polarization vector $\epsilon_{j m}(\sigma) \propto \partial_z Y_{j m}(z,\bar z)$ when, for example, a null shock wave with divergence-free charge current carrying an angular momentum eigenstate $j$ fall into the BH at $v=v_0 >0$, 
\begin{equation}
j^*_v=\frac{Y_{j m}(z,\bar z)}{r^2} \delta (v-v_0)\,. 
\label{inject}
\end{equation}
generating a multipolar radiation field
\begin{equation}
F_\textrm{soft}=\int^{+ \infty}_{- \infty} d v F^{(0)}_{zv}= - \frac{e^2}{j(j+1)}\partial_z Y_{jm}
\label{frad}
\end{equation}
where $F^{(0)}$ indicates the photon term of the field and $Y_{j m}$ are spherical harmonics  \cite{hawk16}. 
A generic non-trivial asymptotic supertranslation field given by an arbitrary combination of spherical harmonics modifies the spacetime in Eq \ref{metricbh} becomes
\cite{compere}
\begin{eqnarray}
&&ds^2 =- \frac{\left( 1-\frac{M}{2\rho_s}\right)^2}{\left( 1+\frac{M}{2\rho_s}\right)^2}dt^2 + \left( 1+\frac{M}{2\rho_s}\right)^4  \nonumber
\\
 &&\Big( d\rho^2 + \left( ((\rho- E)^2 + U)\gamma_{z \bar z}+ \right. \nonumber
\left. (\rho - E) C_{z \bar z} \right)dz d \bar z \Big)
\label{newmetric}
\end{eqnarray}
where $C_{z \bar z} (\theta,\phi) \equiv  -(2D_z D_{\bar z}-\gamma_{z \bar z} D^2)C$ are the components of the supertranslation field $C$, $C_{0,0}$ is the lowest spherical harmonic mode of $C$ with the auxiliary quantities $U(\theta,\phi) \equiv  \frac{1}{8}C_{z \bar z}C^{{z \bar z}}$, $E(\theta,\phi) \equiv  \frac{1}{2}D^2C +C - C_{(0,0)}$, $t_s = t + C_{(0,0)}$
and $\rho_s = \sqrt{(\rho - C + C_{(0,0)})^2 + D_z C D^z C}$. The modified field is revealed by experiments that take into account the effects of the whole sphere around the BH such as an array of rulers around the central object or by deviations from closed null geodesics in strong lensing effects. The simple bending of light of a distant star in a finite solid angular range outside of the supertranslation horizon remains unaffected by supertranslations.

\section{Organized structures of soft photons}

Consider the conserved quantity total angular momentum \textbf{J} of a particle. This can be decomposed in spin angular momentum (SAM), \textbf{S}, and orbital angular momentum (OAM), \textbf{L}.
Whilst the splitting of the total angular momentum in two observables, $\mathbf{J}=\mathbf{L}+\mathbf{S}$, is valid for massive particles, it has not always a precise physical meaning for the EM field and for the photon; in any case SAM and OAM are auxiliary concepts that describe the photon wavefunction properties with respect to rotations and OAM gives the order of the spherical functions involved in the radiation field together with the parity of the photon state, following precise rules for the composition of \textbf{L} and \textbf{S} from the classical field formulation down to the single photon level, where intensity corresponds to the probability of generating a photon in a specific region of spacetime \cite{leach,calvo,tamvic}. 
EM-OAM is currently widely applied in  many research fields \cite{grier} and technology: quantum and classical communications \cite{vaziri,tambu1,willner}, astrophysics \cite{tambu2} and nanotechnology.
Prototype examples of fields carrying specific SAM and OAM eigenvalues are Laguerre-Gaussian (LG) modes that provide an orthogonal fundamental basis to expand any OAM field. They represent cylindrically symmetric structured EM beams that carry $l \hbar$ OAM per photon relative to their symmetry axis with amplitude, in cylindrical coordinates $(r,z,\varphi)$,
\begin{eqnarray} 
\label{eqn:lgmode}
&&u_{lm}^{L-G}(r, \varphi, z) = \sqrt{\frac{2 m!}{\pi (m + l)!}}
 \frac{1}{w(z)} {\left[\frac{r \sqrt{2}}{w(z)}\right]}^l L_m^l \left[\frac{2 r^2}{w^2(z)}\right] 
 \\
&&e^{\frac{-r^2}{w(z)^2} - \frac{\mathrm{i} k r^{2}}{2R(z)}} e^{-\mathrm{i} (2 m + l + 1) \arctan \left(\frac{z}{z_R}\right)} e^{-\mathrm{i} l \varphi } \nonumber
\end{eqnarray}
where $z_R$ is the Rayleigh range of the beam, $w(z)$ the beam waist, $L_m^l (x)$ the associated Laguerre polynomial and $R(z)$ the  curvature radius. The azimuthal and radial indices $l$ and $m$ give the OAM and the number of radial nodes of the associated intensity profile, respectively \cite{thidebook,oambook}. 

All fields carry energy, linear momentum and orbital angular momentum
\cite{Soper:Book:1976}.  The EM field is a vector field and therefore
carries both spin and orbital angular momentum. These observables
are emitted in the form of volumetric densities \cite{review}
and can therefore not be measured at a single point but have to be
integrated/averaged over a finite (possibly very small) volume, for
instance the volume occupied by a sensor.

Any generic field can be written as a superposition of LG modes and decomposed
into multipolar fields, spherical harmonics and in terms of paraxial fields
that depend only on the wavelength \cite{RoseAM}. These properties remain
valid down to the single photon level.  For a simple LG beam with $m=0$
the vector potential $\A$, in terms of multipolar superpositions of
circularly polarized plane waves, is
\begin{equation}
 \A = 2\pi \sum_{j=\|l+p\|}^{\infty} i^j (2j+1)^{1/2} C_{j l p}
  \left[\A_{j\,(l+p)}^{(\mathrm{m})} + i\A_{j\,(l+p)}^{(\mathrm{e})}\right]
\label{potentialA}
\end{equation}
and 
\begin{eqnarray}
&&C_{j l p}=
\sqrt{\frac{(j+p)!(j-p)!(j+l+p)!(j-l-p)!}{(j-\frac{|l|-|l+2p|}{2})!}} \nonumber
 \\
&&k(-)^{j+\frac{l+|l+2p|}{2}}2^{|l|/2+1}w_0^{-2j-1+|l+2p|}L_{j-\frac{|l|+|l+2p|}{2}}^{|l+2p|}\left((w_0/k)^2\right)
\label{gaussian}
\end{eqnarray}
where the vector potential ${\mathbf A}$ is decomposed in the magnetic $(\mathrm{m})$ and electric $(\mathrm{e})$ terms, $k$ is the wavenumber, $w_0$ the waist of the associated Gaussian beam and $L_l^p$ is the associated Laguerre polynomial. The field is composed of multipolar solutions with different total angular momenta, but with the same projection of the angular momentum in the direction $z$, which is $+1$ or $-1$ depending on the polarization of the field and $(j, m = l + p)$, $p$ represents the circular polarization operator and $k$ the wavevector. 
LG modes with OAM $l$ and polarization $p$ present multipolar modes with a fixed component $m=l+p$ along $z$ of the angular momentum \cite{gabi}.

Superpositions of multipolar sources from an arbitrary distribution of charges and currents  $j^*_v$
in a finite source volume $V'$ can generate spatially-structured EM
fields that carry a well-precise combination of spherical harmonics
and OAM with respect to a given point in spacetime.  We consider
fields that are characterized by precise phase structures in space
and OAM values, independently from the radiating system, as angular
momentum and its volumetric density is a property of the field
emitted by every radiation, not the property of the radiator itself
\cite{jackson,thidebook,oambook,thide2}.  We find, numerically and
from simple analytic considerations that when these currents fall
into the static BH, they not only produce structured fields and then
supertranslation states modifying the spacetime, but also induce local
spatial structures in the soft photon hair implant in a neighborhood of
the BH event horizon.

In the numerical simulations we consider, for the sake of simplicity, a distribution of dipolar currents in a set of azimuthally-dephased radiating dipoles on a circle with radius $r_c$  and center $O_c$ that produce a radiating EM field with a well-defined OAM value $l$. The circle is orthogonal to the direction $\mathbf{r}_\bot$ connecting $0_c$ and the center of the BH. We calculate the spatial distribution of the soft photon implant on the event horizon of the BH from Eqs. \ref{inject} and \ref{frad} in a neighborhood of the point $O$.
We assume that the BH radius, $r^*$, is $r^* \gg r_c$ and approximate the metrics near the horizon $r \sim 2M G$ in a small angular region $\theta=0$, where a small neighborhood of an observer can be described in first approximation as a Rindler coordinate system and a set of observers on the direction $r_\bot$ identify, at a first approximation, a class of Rindler observers. For these observers and with the approximations here adopted, the radiating field from the currents is not significantly affected by the free-fall radiating process \cite{rindler,almeida05} neglected in first approximation.

The soft photon hair implant is here induced by four falling dipolar
currents distributed on a circle with radius $r_c=\lambda/10$
have a spatial extension much smaller than the wavelength emitted,
$d=\lambda/100$ and are azimuthally dephased of $\delta \varphi=\pi/4$
to generate an EM field with $l=1$ OAM value calculated with respect
to the center of symmetry of the dipole array with linear polarization
across the $x^1=x$~axis of the new accelerated local frame $(x^i)$. The
upper panels of Fig. 1 show the probability of emission of a photon
in a superposition of OAM states and the spatial phase distribution
of the EM field in a window of $25 \lambda \times 25 \lambda$. The
point of observation is located at $300$ wavelengths distance in the
direction $\mathbf{r}_\bot$.  The lower panels show the probability
of associating a soft photon in the soft photon hair implant and the
corresponding phase profile in the neighborhood of the event horizon,
after neglecting the backreaction of the shell and of the EM field on
the geometry.  All numerical simulations were performed with Matlab
and NEC (Numerical Electromagnetics Code) \cite{nec,nec2,nec3} that
morphed the standard NEC input and output into the correct shape of the
static spacetime geometry here considered through a geometrical optics
transformation from analog gravity \cite{landau,barc05}.

\begin{figure}
\begin{center}
\includegraphics[width=8.5cm]{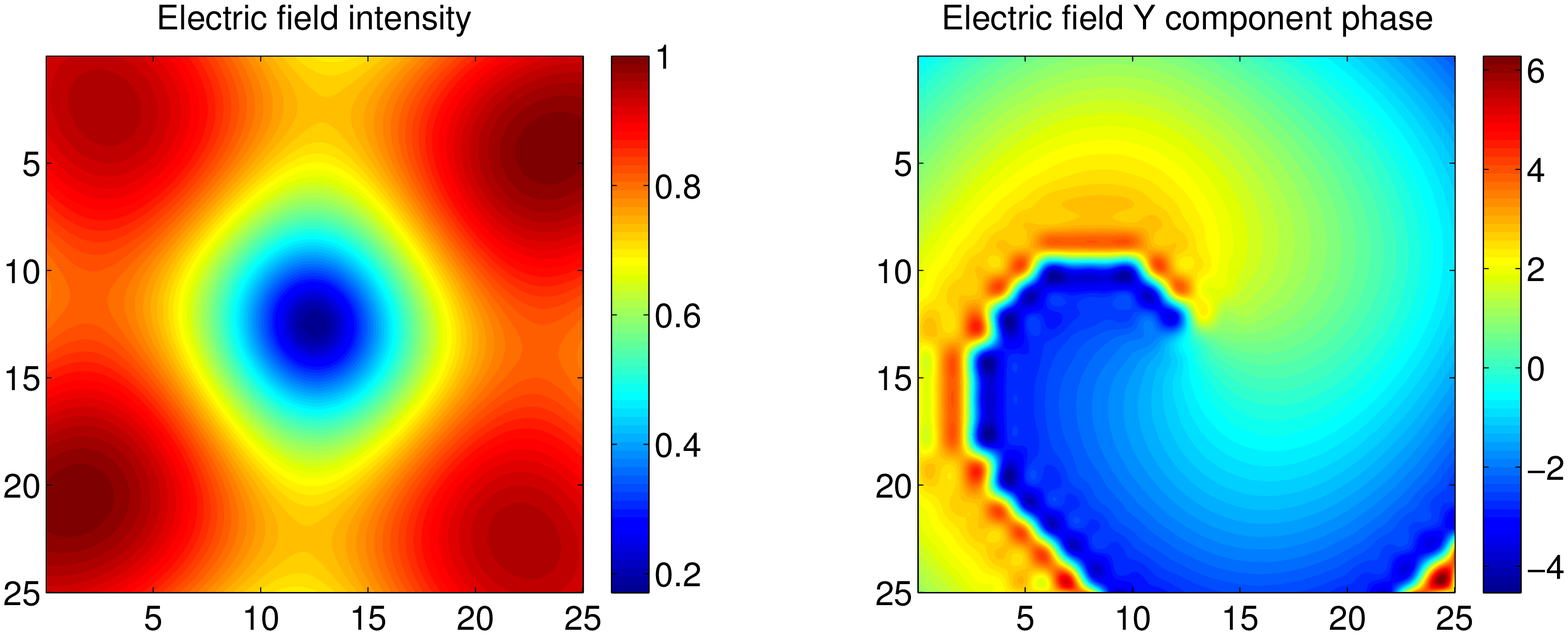}
\includegraphics[width=8.5cm]{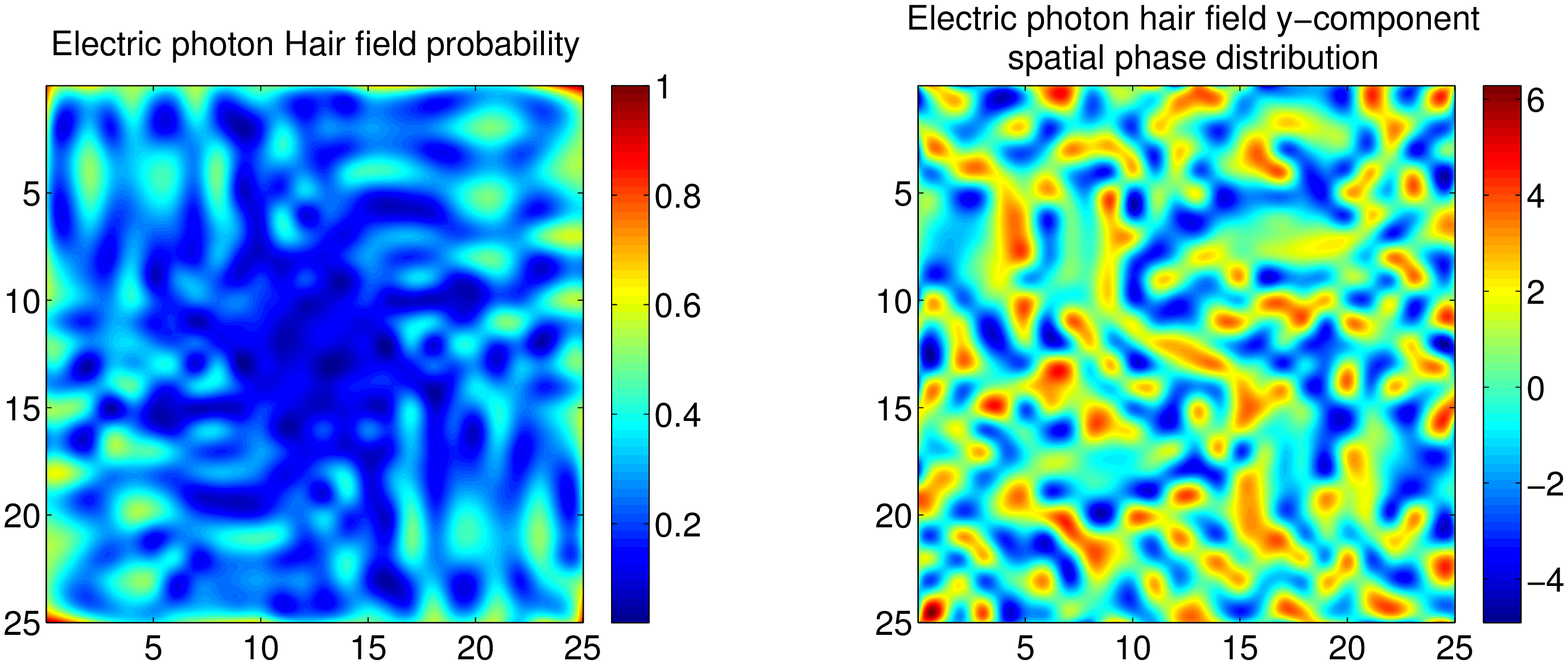}
\end{center}
\caption{Normalized electric field ($E$) probability of photon emission (intensity) and phase distribution of an $l=1$ vortex generated by four dipoles oriented across the y-axis (upper panel). The four dipoles, much smaller than $\lambda$, are distributed on a circle with radius $\lambda /10$. In the lower panel are reported the probability of associating a soft photon hair in the soft photon implant and the spatial phase information generated by the currents.  Units are in $\lambda$.}
\label{fig1}
\end{figure}

\begin{figure}
\begin{center}
\includegraphics[width=8cm,height=4cm]{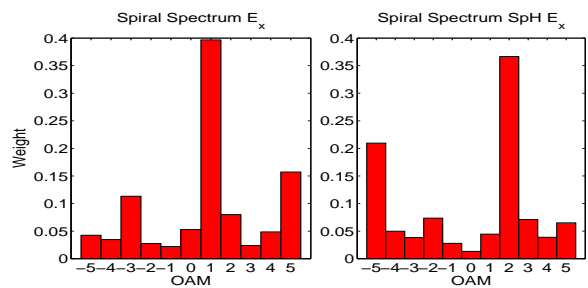}
\end{center}
\caption{Left: spiral spectrum of the radiation field emitted by the dipolar currents shows a dominant contribution of the  $l=1$ vorticity of the source. 
Right: spiral spectrum of the soft photon field is peaked on a dominant value ($l=2$) is larger of one unity with respect to that of the real photon field.}
\label{fig2}
\end{figure}

Owing to the realistic simulation of the four dipoles, the spiral spectrum
of the electric field component along $x$ generated by the dipoles clearly
shows a complex structure of OAM components where the dominant term is
$l=1$ (Fig. 2), as in the spiral spectra \cite{torres} of the EM field
emitted by the dipoles and that of the corresponding soft photon field.
Interestingly, the numerical results show that the spiral spectrum of
the EM field generated by the four falling dipoles, peaked at $l=+1$, has
similarities to that of the soft photon hair implant but for a translation
of the peak to $l=+2$, apparently preserving the spatial information of
the real photon field.  Naively, this effect of spiral spectrum shift
finds an interesting parallelism with the optical experiments involving
OAM beams: the soft photon hair implant $F_\textrm{soft}$ behaves as if
it were the product of the initial radiation field of the current $j^*_v$
after crossing a single-bifurcation fork hologram that shifts the spiral
spectrum of one OAM unit \cite{oambook}.

From an analytical point of view, the shift of the spiral spectrum can
be explained as follows. Consider a set of charges that radiate an EM
field described by a single ideal LG mode with topological charge $l$
as in Eq. \ref{eqn:lgmode}. This carries a finite value of polarization
and OAM, with a precise spatial structure in intensity and phase.
The decomposition in spherical modes of this field puts in immediate
evidence the angular momentum properties  through the eigenfunctions
of the angular momentum operators, being $Y_{jm}$ the eigenfunction
of the operators $J^2$ and $J_z$, where $z$ in this case is $r_\bot$.
By applying Eq. \ref{frad} and the derivation rules in the Riemann sphere
\cite{abramo,arfken}, after some algebra, being
$\frac d{dx} L^l_k = - L_{k-1}^{l+1}$, 
where $l$ and $k$ are two arbitrary indexes, it is clear that the resulting soft photon field presents terms with increased OAM value of exactly one unit,  $l+1$, with respect to the spherical harmonic distribution of the field emitted by the charges of Eqs. \ref{potentialA} and \ref{gaussian}, confirming our numerical findings in the spiral spectra.

All information regarding the spatial distribution of currents are thus encoded on the event horizon and written in the modified metric of Eq. \ref{newmetric}. In this way we provide a novel physical interpretation to the \textit{lush head} of BHs ``soft hair'' that have a modified spacetime with an organized spatial structure on the event horizon, like for ``real'' photons.

{\bf Conclusions.}
Following HPS, Black holes encode the spatial information of infalling currents generating OAM beams in twisted hair implants that present a local a spatial structure on the event horizon. All information is encoded in a new spacetime with supertranslation fields. 
We have shown that we can use the symmetries of EM fields to generate local spatial structures on soft photon implants encoding structured and densely organized information on the event horizon. Thus, BHs can have ``curly'', twisted, structured soft-hair implants where information can be written onto the horizon in an organized way. This procedure can be extended to more complex field configurations expressed in terms of conserved quantities and symmetries inducing additional supertranslation fields.
Interestingly, a coherent and organized superposition of these quantum states corresponds to an organized quantum bits string of information encoded into the event horizon as in an holographic projection.

Multi-dimensional strings of qbits carrying ordered and structured information can excite pixels in an ordered way, creating structures on the event horizon can be built from the invariants and symmetries of the gravitational and EM fields, in a sort of quantum alphabet. This might represent a novel way to input controlled information on BHs that will evolve according to their computational complexity.
Complexity can be encoded through a coherent \cite{dasgupta} superposition and combination of states as a string of a quantum computer \cite{lloyd} where, possibly, a non-thermal spectrum characterizes the passage from macroscopic to microscopic information. 
 This information is thought to be processed by the BH interior as in a quantum computer obeying the rules of quantum computation and complexity; BHs are supposed to be the fastest computers to process complexity in the most optimal way \cite{brown,brown2,hawk14}, encoding information in the structure of spacetime modifying it with the supertranslation fields.
Any BH would resemble a quantum computer with high coherence and this information encoded so far could be located by the soft-hair implant in a neighborhood of the event horizon and in the modified spacetime. 
In any case, because of Shaw theorem that connects information and volume in the phases, any physical system must belong to one of these three classes: systems that preserve information, information dissipating systems and polynomial/exponential information amplifiers \cite{licata1} that for BHs seems to be still an open debate. 

\section*{Acknowledgments}
F.T. gratefully acknowledges the financial support from ZKM and MSC-bw. M.D.L. is supported by ERC Synergy
Grant "BlackHoleCam" Imaging the Event Horizon of Black Holes awarded by the ERC in 2013 (Grant No. 610058).

\end{document}